\documentclass[sigconf]{acmart}

\AtBeginDocument{%
  \providecommand\BibTeX{{%
    \normalfont B\kern-0.5em{\scshape i\kern-0.25em b}\kern-0.8em\TeX}}}

\setcopyright{acmcopyright}
\copyrightyear{2022}
\acmYear{2022}
\acmDOI{XXXXXXX.XXXXXXX}

\acmConference[Conference acronym 'XX]{Make sure to enter the correct
  conference title from your rights confirmation emai}{June 03--05,
  2022}{Woodstock, NY}
\acmPrice{15.00}
\acmISBN{978-1-4503-XXXX-X/18/06}

\acmSubmissionID{1132}

\usepackage{caption}
\usepackage{float}
\usepackage{subcaption}

\usepackage{algorithm}
\usepackage{algorithmic}
\usepackage{threeparttable}
\usepackage{graphicx}
\usepackage{multirow}
\usepackage{pifont}

\begin{document}
\title{Towards Deep Network Steganography: From Networks to Networks}


\author{Guobiao Li}
\email{gbli20@fudan.edu.cn}
\affiliation{%
  \institution{Fudan University}
  \city{}
  \country{}}

\author{Sheng Li}
\email{lisheng@fudan.edu.cn}
\affiliation{%
  \institution{Fudan University}
  \city{}
  \country{}
}

\author{Meiling Li}
\email{mlli20@fudan.edu.cn}
\affiliation{%
 \institution{Fudan University}
 \city{}
 \country{}}

\author{Zhenxing Qian}
\email{zxqian@fudan.edu.cn}
\affiliation{%
  \institution{Fudan University}
  \city{}
  \country{}}

\author{Xinpeng Zhang}
\email{zhangxinpeng@fudan.edu.cn}
\affiliation{%
  \institution{Fudan University}
  \city{}
  \country{}
  }


\renewcommand{\shortauthors}{Li et al.}

\begin{abstract}
With the widespread applications of the deep neural network (DNN), how to covertly transmit the DNN models in public channels brings us the attention, especially for those trained for secret-learning tasks. In this paper, we propose deep network steganography for the covert communication of DNN models. Unlike the existing steganography schemes which focus on the subtle modification of the cover data to accommodate the secrets, our scheme is learning task oriented, where the learning task of the secret DNN model (termed as secret-learning task) is disguised into another ordinary learning task conducted in a stego DNN model (termed as stego-learning task). To this end, we propose a gradient-based filter insertion scheme to insert interference filters into the important positions in the secret DNN model to form a stego DNN model. These positions are then embedded into the stego DNN model using a key by side information hiding. Finally, we activate the interference filters by a partial optimization strategy, such that the generated stego DNN model works on the stego-learning task. We conduct the experiments on both the intra-task steganography and inter-task steganography (i.e., the secret and stego-learning tasks belong to the same and different categories), both of which demonstrate the effectiveness of our proposed method for covert communication of DNN models.


\end{abstract}

\ccsdesc[500]{Security and privacy}


\keywords{deep neural networks, information hiding, steganography}


\maketitle

\section{Introduction}

\begin{figure}[ht]
	\centering
	\includegraphics[width=1\linewidth]{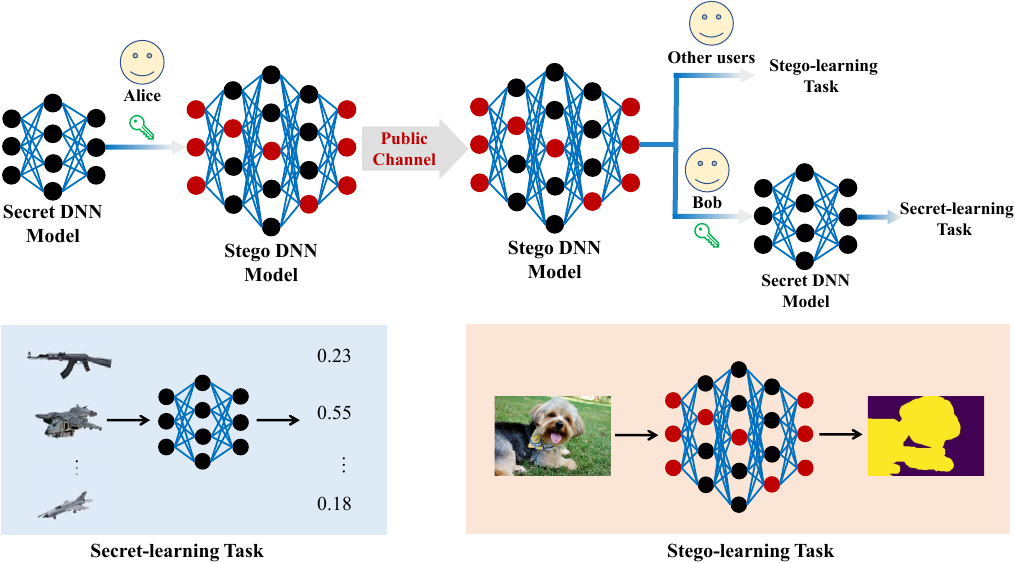}
	\caption{Illustration of the application scenario of the proposed deep network steganography. \label{fig:scenario}} 
\end{figure}

Steganography is a technique of hiding secret messages into digital cover data imperceptibly for covert communication. Various types of cover data have been considered in literature on steganography, including images \cite{8789545, 9124671, liao2019new}, text \cite{krishnan2017overview}, video \cite{mstafa2020new}, audio \cite{gopalan2003audio} or 3D Mesh \cite{zhou2019distortion}. Most of the existing steganographic schemes focus on designing hand-crafted algorithms to modify the cover data subtly for data embedding, which suffer from limited data embedding capacity (termed as capacity for short). To increase the capacity, some deep neural network (DNN)-based steganographic schemes have been proposed recently \cite{Baluja2019image, lu2021large, jing2021hinet}, which are able to hide secret images into the cover images. However, the secret images can not be losslessly recovered during data extraction.

Nowadays, deep learning techniques have been applied in various applications such as computer vision \cite{krizhevsky2012imagenet, he2016deep}, natural language processing \cite{goldberg2017neural, devlin2018bert}, information security \cite{zhu2018hidden, Baluja2019image}, etc. How to transmit the DNN models covertly in public channels is worthy of investigation, especially for those trained for secret tasks. One possible solution is to treat the DNN models as ordinary data and hide them into popular cover data such as images or videos using the existing steganographic schemes. In general, the size of a DNN model is relatively large in terms of secret messages. Therefore, using traditional hand-crafted steganographic schemes would result in the need of a large amount of cover data due to the limited capacity. This would cause a lot of communication burden. While the existing high capacity DNN-based steganographic schemes are proposed for hiding secret images with meaningful content, it is difficult to apply them on secret DNN models for lossless data extraction.

To deal with this issue, we propose deep network steganography in this paper, where a stego DNN model (i.e., the DNN with hidden data) is constructed directly from the secret DNN model according to a key, as shown in Figure \ref{fig:scenario}. The stego DNN model performs a different learning task from the secret DNN model. Receivers with the correct key can extract the secret DNN model losslessly from the stego DNN model for the secret-learning task. Other users (or people monitoring the channel) may not notice the existence of secret DNN model, who could only use the stego DNN model to conduct the stego-learning task. Compared with the existing steganographic schemes, such a network to network steganography is tailored for covert communication of DNN model, which has the following advantages.
\begin{itemize}
  \item [1)] The communication burden hardly increased before and after steganography, and there is no need to collect a large amount of cover data to accommodate the secret DNN model.

  \item [2)] The behaviour of covert communication is well protected by placing another functional DNN with a stego-learning task. The stego DNN model could be placed in the DNN model repository that is public available (such as the github model repository) without changing the nature of DNN models.
\end{itemize}

The main contributions of this paper are summarized below.

\begin{itemize}
  \item [1)] We propose to construct a stego DNN model by adding interference filters into the secret DNN model, where a partial optimization strategy (POS) is proposed for lossless recovery of the secret DNN model.
  \item [2)] We propose a gradient-based filter insertion (GFI) strategy for adding the interference filters, which guarantees the construction of a stego DNN model with good performance using a few interference filters.
  \item [3)] We design two statistical loss terms for training a stego DNN model with  statistical undetectability.  
  \item [4)] We perform extensive experiments among different learning tasks such as image classification, image segmentation and image denoising, all of which demonstrate the effectiveness of our proposed method for network steganography.
\end{itemize}

\vspace{0.3mm}
\section{Related Works}
\label{section:related}

\subsection{Steganography}
According to whether DNN is used for data embedding or extraction, steganography schemes can be mainly divided into traditional steganography and DNN-based steganography. Liao \textit{et al.} \cite{liao2019new} conduct data embedding on color images and assign different payloads to each channel according to the proposed amplification channel modification probability strategy. Lu \textit{et al.} \cite{8789545} propose to convey secret data through halftoned image, where the histogram of the pixels is slightly modified to accommodate the secret message. Mstafa \textit{et al.} \cite{mstafa2020new} choose to hide confidential data inside the corner points within the cover video frames identified by Shi-Tomasi algorithm. Zhou \textit{et al.} \cite{zhou2019distortion} propose to embed secret information into 3D meshes through geometric modifications, where the importance of different regions in the 3D meshes is measured for adaptive data embedding. To ensure the undetectability, the capacity of these schemes are limited. Take the image steganography as an example, each pixel in a grayscale image (8 bits) could accommodate 1 bit of secret message at most for high undetectability.

Most of the DNN steganographic schemes are designed for image steganography. Some works utilize DNN as a preprocessing strategy before data embedding. Tang \textit{et al.} \cite{tang2017automatic} propose to use the generative adversarial network (GAN) to automatically learn a distortion function for adaptive data embedding in images, which consists of a steganographic generative sub-network and a steganalytic discriminative sub-network. This is further improved in \cite{tang2020automatic} by incorporating a reinforcement learning framework to learn an optimized embedding cost for each pixel. With the distortion function or embedding cost available, these two schemes still use handcrafted algorithms for data embedding. Thus, their capacity are also limited.

Recently, more and more DNN-based steganographic schemes are proposed without any handcrafted processes, which usually involve training deep encoder-decoder networks to hide and recover secret data. Zhu \textit{et al.} \cite{zhu2018hidden} conducte the early research on such technique, where the secrets are embedded into a cover image using an end-to-end learnable DNN. Encouraged by \cite{zhu2018hidden}, some researchers propose to hide secret images into cover images for high capacity data embedding 
\cite{Baluja2019image, lu2021large, jing2021hinet}. The concept of hiding images into images is initially proposed by Baluja \textit{et al.} \cite{Baluja2019image}, where a preparation network is adopted to extract the features of secret image, and then a hiding network encodes the features and cover image for generating the stego-image. The stego-image is fed into an image decoding network eventually to extract the secret image. The above three networks are jointly learnt for minimized extraction errors and image distortion. 
Similarly, the work \cite{lu2021large} makes use of the invertible neural network for image encoding and decoding, where the number of the channels in the secret image branch can be increased, \textit{i.e.} concatenating multiple secret images along the channel axes, to remarkably improve the capacity. 
Jing \textit{et al.} \cite{jing2021hinet} hide the secret image in wavelet domain of cover image by the proposed low-frequency wavelet loss, which significantly improves the hiding undetectability.
Despite the advantage, the works in \cite{Baluja2019image, lu2021large, jing2021hinet} are designed to hide structured images instead of random bits. In addition, the secret images can not be losslessly recovered in image decoding.

\begin{figure*}[htbp]
	\centering
	\includegraphics[width=0.80\linewidth]{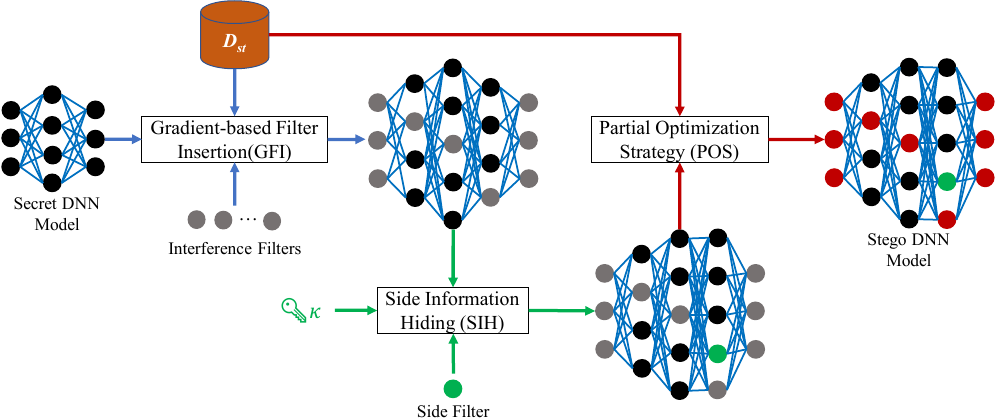}	\caption{The proposed method for DNN model embedding.}
	\label{fig:overflow}
\end{figure*}


\subsection{Watermarking}
Watermarking tries to embed ownership information or authenticated information into the cover data for copyright protection or tampering detection. Researchers have devoted efforts in developing watermarking schemes for DNN models including robust DNN watermarking and fragile DNN watermarking.

The robust DNN watermarking schemes are designed for protecting DNN models against unauthorized usages \cite{uchida2017embedding, darvish2019deepsigns, fan2021deepip}. In these schemes, the ownership of the DNN model is embedded into the model parameters by fine-tuning, where additional regularization terms are incorporated in the loss functions for watermark embedding. The fragile DNN watermarking schemes \cite{guan2020reversible, abuadbba2021deepisign} are proposed to detect whether the DNN models are tampered such as poisoned or fine-tuned, where some authenticated information is embedded by modifying the DNN model parameters after training. Guan \textit{et al.} \cite{guan2020reversible}  propose to embed the hash of the original DNN into the DNN parameters by a histogram shift strategy, where the parameters are transformed into integers and slightly modified to accommodate the secret bits. Abuadbba \textit{et al.} \cite{abuadbba2021deepisign} propose to embed the secret message and its hash in the frequency domain of the DNN model parameters, where the less significant coefficients are replaced with hidden data. These schemes do not take the undetectability into account, which are only able to insert a few bits for copy right protection or tampering detection.

Despite the progress in the area of information hiding, little attention has been paid to the problem of covert communication of DNN models. In this paper, we try to tackle this problem by a network to network steganographic mechanism, which is able to recover the secret model losslessly with good performance in capacity and undetectability.

\section{Problem Formulation}
\label{sec:goals}

Let's denote the secret DNN model as $F_\theta(\cdot)$ and the constructed stego DNN model as $G_{\delta}(\cdot)$, respectively. During the model embedding, the sender first selects an appropriate learning task (i.e., the stego-learning task) for $G_{\delta}(\cdot)$. Then, he constructs $G_{\delta}(\cdot)$ by adding some interference parameters (say $\theta_a$) to $F_\theta(\cdot)$ using a key $\kappa$ with the final layer adapted to the stego-learning task. Next, the sender tunes the interference parameters in $G_{\delta}(\cdot)$ using a stego-dataset to build a functional stego DNN model. Upon receiving the stego DNN model, the receiver extracts the original parameters of $F_{\theta}(\cdot)$ from $G_{\delta}(\cdot)$ by using the same key $\kappa$. The aforementioned procedure can be formulated below for the sender and receiver.
\begin{equation}
\left\{ \begin{array}{l}
   Sender: G_{\delta}(\cdot) = Emb \left(F_{\theta}(\cdot), \theta_a, \kappa \right)\\
   Receiver: F_{\theta}(\cdot) = Ex \left(G_{\delta}(\cdot), \kappa \right)\\
\end{array}\right. ,
\label{eq1}
\end{equation}

\noindent where $Emb\left(\cdot\right)$ and $Ex\left(\cdot\right)$ refer to model embedding and extraction, respectively.

The design of $Emb\left(\cdot\right)$ and $Ex\left(\cdot\right)$ should satisfy the following properties for covert communication.
\begin{itemize}
  \item [1)] \textbf{Recoverability}: the secret DNN model should be fully recovered in model extraction.
  \item [2)] \textbf{Fidelity}: the performance of the stego DNN model should be reasonably good for the stego-learning task.
  \item [3)] \textbf{Capacity}: the size of the stego DNN model should not expand too much compared with the secret DNN model for efficient transmission. 
  \item [4)] \textbf{Undetectability}: the statistical features of the stego DNN model should be close to those of the ordinary models to avoid being detected.
\end{itemize}



\begin{figure*}[htbp]
	\centering
	\includegraphics[width=0.82\linewidth]{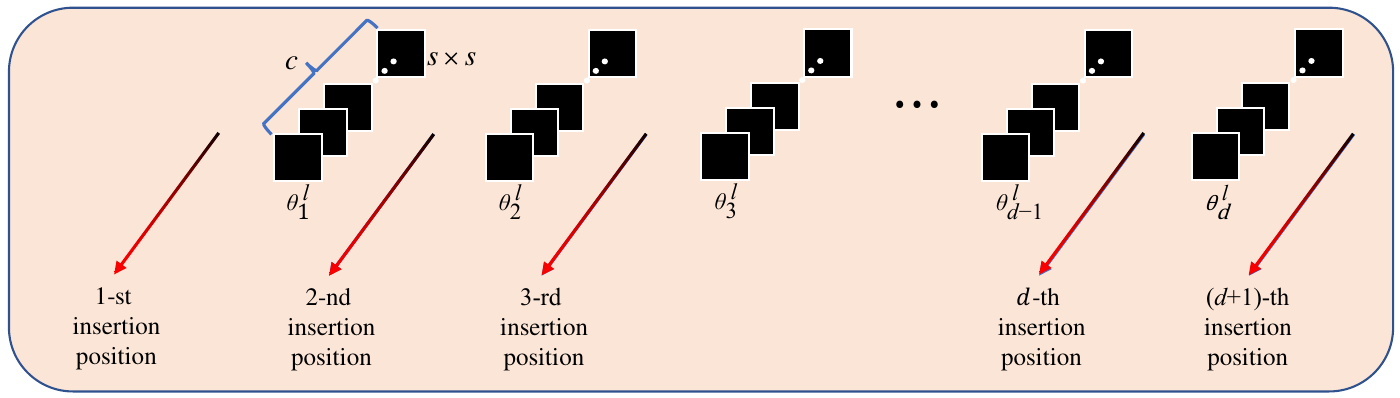}
	\caption{Insertion positions in the \textit{l}-th convolutional layer.}
	\label{fig:insertion_position}
\end{figure*}

\section{The Proposed Method}

Given a well trained secret DNN model $F_\theta(\cdot)$, we conduct model embedding in three sequential steps as shown in Figure~\ref{fig:overflow}, including 1) gradient-based filter insertion (GFI); 2) side information hiding (SIH); and 3) stego DNN model training via a partial optimization strategy (POS). The GFI establishes the architecture of the stego DNN model $G_{\delta}(\cdot)$ from $F_\theta(\cdot)$ by inserting interference parameters (in terms of interference filters) according to the gradient of the filters in $F_\theta(\cdot)$. The SIH conceals the positions of the interference filters into $G_{\delta}(\cdot)$. The POS trains $G_{\delta}(\cdot)$ by tuning the interference parameters according to a stego dataset for the stego-learning task. In what follows, we elaborate each step in detail.

\vspace{-2mm}
\subsection{Gradient-based filter insertion}
\label{sec:gps}

Let's assume $F_{\theta}(\cdot)$ has $L$ convolutional layers. We denote the $l$-th convolutional layer as $\theta^l \in \mathbb{R}^{d \times c \times s \times s}$, where $d$ is the number of filters, $c$ is the number of channels of each filter, and $s\times s$ is the kernel size. Each filter can be represented as $\theta_i^l \in \mathbb{R}^{c \times s \times s}$ with $i$ being the index. The basic concept of our model embedding is to insert interference filters into the convolutional layers of $F_{\theta}(\cdot)$ to form a stego DNN model. For the $l$-th convolutional layer, there are $(d+1)$ positions that could be selected to insert the interference filters, as shown in Figure \ref{fig:insertion_position}. For simplicity, We denote such positions as the insertion positions. A straightforward way is to randomly select the insertion positions for filter insertion. However, such a strategy may not be efficient because the randomly selected positions may not be sensitive for building the stego DNN model. We may have to insert a lot of interference filters in order to achieve high fidelity of the stego DNN model. To deal with this issue, we propose here a gradient-based filter insertion scheme which performs filter insertion in the most important positions for the stego-learning task.

First of all, we measure the importance of each insertion position below. We obtain the gradient of $\theta^l$ by feeding the stego dataset $D_{st}$ into $F_{\theta}(\cdot)$, which is computed as
\begin{equation}
\hat{\theta}^l = \sum_{(x, y) \in D_{st}} | \nabla_{\theta^l} \mathcal{L}_{st}(x, y) |,
\label{eq2}
\end{equation}
where $x$, $y$ refer to the samples and the corresponding labels in $D_{st}$. $\mathcal{L}_{st}$ is a loss function for the stego-learning task. For each filter, we measure its importance as
\begin{equation}
w^l_{i} = mean (\hat{\theta}^{l}_{i}),
\end{equation}
where $\hat{\theta}^{l}_{i}$ refers to the gradients computed from the $i$-th filter and $mean(\cdot)$ is a function to get the mean value of a three dimensional tensor. The importance of each insertion position is then measured according to the importance of its two neighboring filters, which is computed as
\begin{equation}
    p^l_j = \frac{1}{2} (w^l_{j-1} + w^l_{j}),
\label{eq4}
\end{equation}
where $j \in [2, d]$ is the index of the position. Here, we set $p^l_1$=$w^l_1$ and $p^l_{d+1}$=$w^l_d$ because the 1st and $(d+1)$-th insertion position only has one neighboring filter.

We then select the top-$N$ most important insertion positions among all the insertion positions of the convolutional layers. Starting from the most important position, we gradually insert the interference filters and update the network architecture of $F_{\theta}(\cdot)$. For each selected position, the architecture of $F_{\theta}(\cdot)$ is updated below: 1) insert an interference filter with random parameters sized $c'\times s' \times s'$, where $c'$ is the channel number in the current layer and  $s' \times s'$ is the size of the kernel; 2) update the next convolutional layer by adding one interference channel with random parameters in all the filters of this layer. The aforementioned process is repeated until all the selected positions are processed and a stego DNN model $G_{\delta}(\cdot)$ is constructed accordingly. 



Finally, we record the locations of the interference filters in $G_{\delta}(\cdot)$ into $L$ binary streams $\mathbb{B}=\{b_1, b_2, \cdots, b_L\}$. Each layer corresponds a $d'$-bits binary stream $b_i$ with $d'$ being the number of filters in this layer, and each bit corresponds to a filter with $0$ being the interference filter and $1$ being the non-interference filter. Algorithm \ref{alg:GPS} gives the pseudo code of our proposed gradient-based filter insertion, where $Top_N(\cdot)$ and $InsertFilters(\cdot)$ refers to the operation of the insertion positions selection and filter insertion, and $\delta^l_k$ is the $k$-th filter in the $l$-th layer of $G_{\delta}(\cdot)$.



\subsection{Side Information Hiding}
The purpose of side information hiding is to conceal the locations of interference filters (i.e., $\mathbb{B}$) into the stego DNN model $G_{\delta}(\cdot)$, which is performed by inserting an additional filter (termed as the side filter) in $G_{\delta}(\cdot)$ with $\mathbb{B}$ hidden.

To conduct side information hiding, we use a key $\kappa$ to determine the insertion position of the side filter in $G_{\delta}(\cdot)$, where its parameters are randomly set. With the side filter inserted, we encrypt $\mathbb{B}$ to $\mathbb{B}'$ using $\kappa$ and embed $\mathbb{B}'$ into the parameters of the side filter. The data embedding algorithm here is motivated by the work proposed in \cite{guan2020reversible}. In particular, we transform the parameters in the side filter into integers, whose least significant bits are replaced with $\mathbb{B}'$ for data embedding. The parameters with hidden data are then inversely transformed into the floating numbers to complete the process.

\begin{algorithm}[htb]
\caption{Gradient-based filter insertion.}
\label{alg:GPS}
\begin{algorithmic}[1] 
\REQUIRE ~~\\ 
    Secret DNN model parameters: $\theta$ ($L$ layers)\\
    Interference filters: $\theta_a$ (quantity: $n$)\\
    Stego dataset: $D_{st}$\\
\ENSURE ~~\\ 
    Stego DNN model: $G_{\delta}(\cdot)$ \\
    Record of the interference positions: $\mathbb{B}$
    \STATE $\hat{\theta} \gets 0$, $w\gets 0$, $p \gets 0$, $\mathbb{B} \gets \emptyset$ \\
    \STATE \textbf{for} $(x,y)$ in $D_{st}$ \textbf{do}
    \STATE $\qquad \hat{\theta} \gets \hat{\theta} + |\nabla_{\theta} \mathcal{L}_{st}(\theta, x, y)$|
    \STATE \textbf{end for}

    \STATE \textbf{for} $l$ in $L$ \textbf{do}
    \STATE $\qquad$ \textbf{for} $i$ in $d$ \textbf{do}
    \STATE $\qquad$ $\qquad$ $w^l_{i} \gets mean (\hat{\theta}^{l}_{i})$
    \STATE $\qquad$ \textbf{end for}
    \STATE \textbf{end for}

    \STATE \textbf{for} $l$ in $L$ \textbf{do}
    \STATE $\qquad$ $p^l_{1} \gets w^l_{1}$
    \STATE $\qquad$ \textbf{for} $j$ in $d-1$ \textbf{do}
    \STATE $\qquad$ $\qquad$ $p^l_{j} \gets \frac{1}{2} (w^l_{j-1} + w^l_{j})$
    \STATE $\qquad$ \textbf{end for}
    \STATE $\qquad$ $p^l_{d+1} \gets w^l_d$
    \STATE \textbf{end for}
    \STATE $Positions \gets Top_{N}(n)$
    \STATE $G_{\delta}(\cdot) \gets InsertFilters(\theta, \theta_a, Positions)$ 
    
    \STATE \textbf{for} $l$ in $L$ \textbf{do}
    \STATE $\qquad$ $b_l \gets null$
    \STATE $\qquad$ \textbf{for} $k$ in $d^{'}$ \textbf{do} \; 
    \STATE $\qquad$ $\qquad$ $b_l \gets b_l \cup 1$, \ \textbf{if} $\delta^l_{k} \in$ Non-interference filters
    \STATE $\qquad$ $\qquad$ $b_l \gets b_l \cup 0$, \ \textbf{if} $\delta^l_{k} \in$ Interference filters
    \STATE $\qquad$ \textbf{end for}
    \STATE $\qquad$ $\mathbb{B} \gets \mathbb{B} \cup b_l$
    \STATE \textbf{end for}
    
\RETURN $G_{\delta}(\cdot)$, $\mathbb{B}$; 
\end{algorithmic}
\end{algorithm}

\vspace{2mm}
\subsection{Stego Network Training}
In this section, we explain how we train $G_{\delta}(\cdot)$ to obtain a functional stego DNN model for the stego-learning task.

\textbf{Loss function} The training is conducted on a stego-dataset for learning the stego-task and the loss function is defined below.
\begin{equation}
    \mathcal{L}_{all} = \mathcal{L}_{st} + \alpha \mathcal{L}_\mu + \beta \mathcal{L}_{\sigma},
\label{l_t}
\end{equation}
where $\alpha$ and $\beta$ are weights to balance different losses, $\mathcal{L}_\mu$ and $\mathcal{L}_{\sigma}$ are two statistical loss functions for undetectability, where
\begin{equation}
\left\{ \begin{array}{l}
  \mathcal{L}_\mu = \sum_{l=1}^L \| \delta_{\mu}^{l} - \gamma_{\mu}^{l} \|_2 \\
   \mathcal{L}_\sigma = \sum_{l=1}^L \| \delta_{\sigma}^{l} - \gamma_{\sigma}^{l} \|_2 \\
\end{array}\right. ,
\label{l_d}
\end{equation}
where $\| \|_2$ is the L2 norm, $\delta_{\mu}^{l}$ and $\delta_{\sigma}^{l}$ are the mean and standard deviation of the parameters in the $l$-th convolutional layer of the stego DNN model $G_{\delta}(\cdot)$, while $\gamma_{\mu}^{l}$ and $\gamma_{\sigma}^{l}$ are the mean and standard deviation in the $l$-th convolutional layer of the ordinary clean DNN model $G_{\gamma}(\cdot)$ which has the same structure as $G_{\delta}(\cdot)$, but without the secret model embedded. 

Here, we consider only the mean and standard deviation for statistical undetectability due to the following reason. Researchers have indicated that the distribution of the DNN parameters tend to be Gaussian after training \cite{tian2021pss, huang2021rethinking}, and using mean and standard deviation is sufficient to describe a Gaussian distribution.

\textbf{Partial Optimization Strategy}~There are three different types of parameters in the stego DNN $G_{\delta}(\cdot)$, including the original parameters in the secret DNN model, the parameters in the side filter, and the rest which are the interference parameters. To make sure that $F_{\theta}(\cdot)$ can be extracted from $G_{\delta}(\cdot)$ losslessly, we can not modify the first two types of the parameters, only the interference parameters can be tuned to achieve high fidelity for $G_{\delta}(\cdot)$. To this end, we introduce a mask $\mathcal{M}$ for partial optimization of $G_{\delta}(\cdot)$. $\mathcal{M}$ is a binary mask with the same size as that of the parameters of $G_{\delta}(\cdot)$ (say $\delta$), where
\begin{equation}
    \mathcal{M}[k] = \left\{\begin{array}{lc}
    1,\quad & \delta[k] \in \Omega  \\
    0,\quad & else \end{array}\right. ,
\label{mask}
\end{equation}
where $\Omega$ is the interference parameter set in $\delta$ and $\delta[k]$ is the $k$-th parameter in $\delta$.

Let $\lambda$ be the learning rate and $\odot$ denotes the element-wise product, $G_{\delta}(\cdot)$ is updated below using gradient descent.
\begin{equation}
    \delta = \delta - \lambda \ \mathcal{M} \odot \ \nabla_{\delta} \mathcal{L}_{all}.
\label{PSS}
\end{equation}

\subsection{Model extraction}
In model extraction, the receiver uses the key $\kappa$ to identify the side filter in the stego DNN model and transforms its parameters into integers. The encrypted side information $\mathbb{B}'$ can be extracted according to the least significant bits of these integers. $\mathbb{B}'$ is then decrypted using $\kappa$ to recover the locations of the interference filters $\mathbb{B}$. Next, the side filter is removed from the stego DNN model, after which the interference filters can be identified based on the location information $\mathbb{B}$. As such, the secret DNN model is extracted accordingly. Since we do not touch the original parameters in the secret DNN model when constructing and training the stego DNN model, the secret DNN model can be losslessly recovered in model extraction.

\begin{figure*}[htbp]
	\centering
	\includegraphics[width=0.84\linewidth]{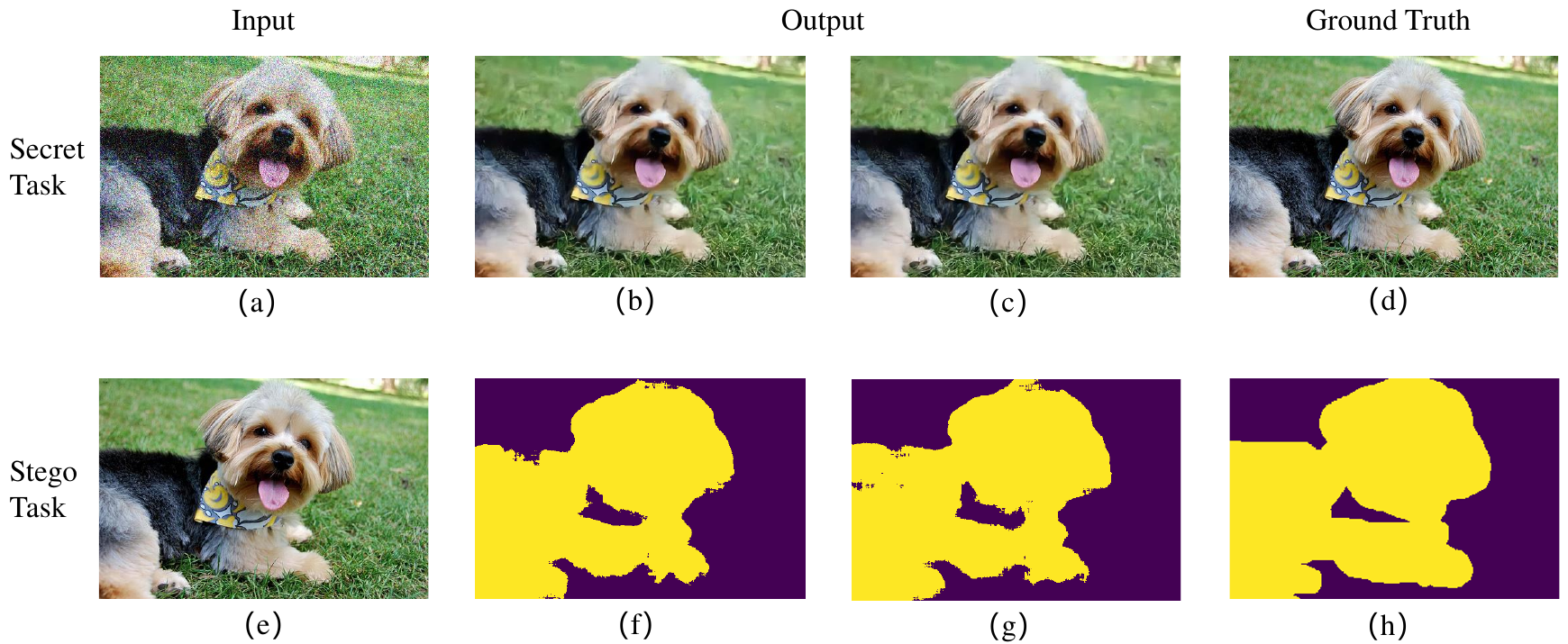}
	\caption{The performance of the secret DNN model and the corresponding stego DNN model on inter-task steganography. (a) A pet image from the Oxford-pet dataset with Gaussian noise (PSNR=14.86dB); (b) the output of the original secret DNN model (PSNR=25.87dB); (c) the output of the recovered secret DNN model (PSNR=25.87dB); (d) the ground truth of (a); (e) a pet image from the Oxford-pet dataset; (f) the output of the stego DNN model (mIOU=89.17\%); (g) the output of the clean DNN model (mIOU=90.17\%); (h) the ground truth of (e). }
	\label{fig:visual_DnCNN_noise}
\end{figure*}

\vspace{2mm}
\section{Experiments}
\subsection{Settings}

We conduct two types of experiments to evaluate the performance of our proposed method according to different steganography scenarios, including
\begin{itemize}
  \item [1)] Intra-task steganography: the secret DNN model and the stego DNN model perform the same type of learning tasks. 
  \item [2)] Inter-task steganography: the secret DNN model and the stego DNN model perform different types of learning tasks.
\end{itemize}

For intra-task steganography, both the secret and stego DNN models perform image classification tasks. In particular, we choose VGG11 \cite{simonyan2015vgg} as the secret DNN model, which is adopted to construct a stego DNN model. Both the two models are trained for image classification tasks. The Fashion-MNIST \cite{xiao2017fashion} and CIFAR10 \cite{krizhevsky2009learning} datasets are used to train the secret DNN model and the stego DNN model, respectively. As such, CIFAR10 acts as $D_{st}$ for DNN steganography. For both datasets, we use their default dataset partition for training and testing. The classification accuracy is the performance indicator here, which is termed as ACC in the following discussions. 
 
For inter-task steganography, we decide to hide an image denoising model into an image segmentation model. We download a well-trained DnCNN \cite{zhang2017beyond} model as secret DNN model \footnote{https://github.com/cszn/KAIR/tree/master/model\_zoo}. We adopt the benchmark image segmentation dataset Oxford-Pet \cite{parkhi2012cats} as $D_{st}$ to train and test the stego DNN model constructed from the secret DNN model. Specifically, we randomly separate the Oxford-Pet dataset into three parts, including 6000 images for training, 1282 images for validating and 100 images for testing. We evaluate the performance of stego DNN model on the Oxford-Pet testset and take the Mean Intersection over Union (mIOU) as the performance indicator. When evaluating the secret DNN model (i.e., the DnCNN model), we pre-process the images in the Oxford-Pet test set by adding Gaussian noise with a noise level of 50. Then, we feed these noisy images into the secret DNN model and compute the peak single noise ratio (PSNR) between the model predictions and the original test images to evaluate the performance.  

For both steganographic scenarios, we select top $30\%$ most important insertion positions for filter insertion, where $N$ is 825 for intra-task steganography and 384 for inter-task steganography. For training the above models, we use Adam \cite{kingma2014adam} optimizer with learning rate $\lambda=0.001$ and set hyperparameters $\alpha=20.0$, $\beta=1.0$. All our experiments are conducted on Ubuntu 18.04 system with four NVIDIA RTX 2080 Ti GPUs.

\begin{table}[htbp]
	\centering
	\caption{The recoverability of the proposed method.}
	\label{tab:recoverability}
	\resizebox{0.8\linewidth}{!}{%
	\begin{threeparttable}
		\begin{tabular}{l|cc|cc}
			\hline
			\multirow{2}{*}{Model} & \multicolumn{2}{c|}{Intra-task} & \multicolumn{2}{c}{Inter-task}     \\
			& BER(\%)  & ACC (\%)  & BER(\%)  & PSNR (db) \\
			\hline
			$F_\theta(\cdot)$       & -  & 93.67   & - & 27.42   \\
			$F_{\theta^{'}}(\cdot)$ & 0.00  & 93.67   & 0.00 & 27.42   \\
		     \hline 
		\end{tabular}%
	\end{threeparttable}}
\end{table}

\begin{table}[htbp]
	\centering
	\caption{The fidelity of the proposed method.}
	\label{tab:reliablity}
	\resizebox{0.5\linewidth}{!}{%
	\begin{threeparttable}
		\begin{tabular}{l|c|c}
			\hline
			\multirow{2}{*}{Model} & \multicolumn{1}{c|}{Intra-task} & \multicolumn{1}{c}{Inter-task}     \\
			& ACC (\%)  & mIOU (\%)  \\
			\hline
			$G_{\delta}(\cdot)$ & 90.52  & 75.43    \\
			$G_{\gamma}(\cdot)$ & 91.45  & 76.67  \\
		     \hline 
		\end{tabular}%
	\end{threeparttable}}
\end{table}

\subsection{Performance}
As what have been discussed in Section \ref{sec:goals}, we evaluate the performance of the proposed method in terms of recoverability, fidelity, capacity and undetectability.
\vspace{3mm}
\subsubsection{\textbf{Recoverability}}~
We compute the bit error rate (BER) between the original secret DNN model $F_{\theta}(\cdot)$ and the recovered secret DNN model (say $F_{\theta^{'}}(\cdot)$) which is extracted from the stego DNN model. Table \ref{tab:recoverability} gives the BER for different steganography scenarios as well as the performanc of the secret DNN model. It can be seen that the secret DNN model can be extracted losslessly. Therefore, its performance is not affected at all in our model steganography.

\subsubsection{\textbf{Fidelity}}
To evaluate the fidelity of the stego DNN model $G_{\delta}(\cdot)$, we train a clean model $G_{\gamma}(\cdot)$ which has the same architecture as $G_{\delta}(\cdot)$. The training of the $G_{\gamma}(\cdot)$ is conducted on the same training set as that is used for training $G_{\delta}(\cdot)$, where all the parameters in $G_{\gamma}(\cdot)$ are tuned to achieve the best performance. Furthermore, we use the same test set to evaluate the performance of the two models.

Table \ref{tab:reliablity} shows the performance of the stego DNN model and the clean model for both intra-task steganography and inter-task steganography . For intra-task steganography, it can be seen that the ACC of the stego DNN model on CIFAR10 is reduced with less than $1\%$ after the model embedding. The results of the inter-task steganography are similar to that of intra-task steganography. In particular, after model embedding, the mIOU of the stego DNN model is slightly degraded when compared with the corresponding clean model $G_{\gamma}(\cdot)$ on the Oxford-Pet dataset. These results indicate that the fidelity of our stego DNN model is favorable, which has little performance degradation caused by model embedding. Figure \ref{fig:visual_DnCNN_noise} visualizes the results of the secret DNN model and the stego DNN model for different learning tasks. 

\begin{figure*}[htbp]
	\centering
	\begin{minipage}{0.49\linewidth}
		\centering
		\includegraphics[width=0.85\linewidth]{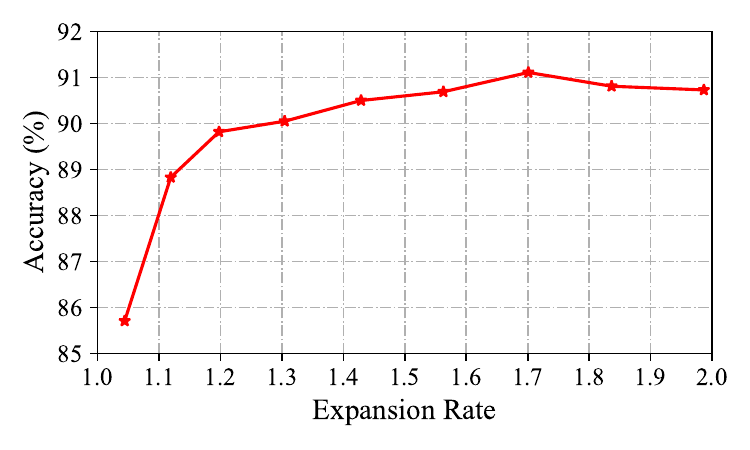}
	\end{minipage}
	\begin{minipage}{0.49\linewidth}
		\centering
		\includegraphics[width=0.85\linewidth]{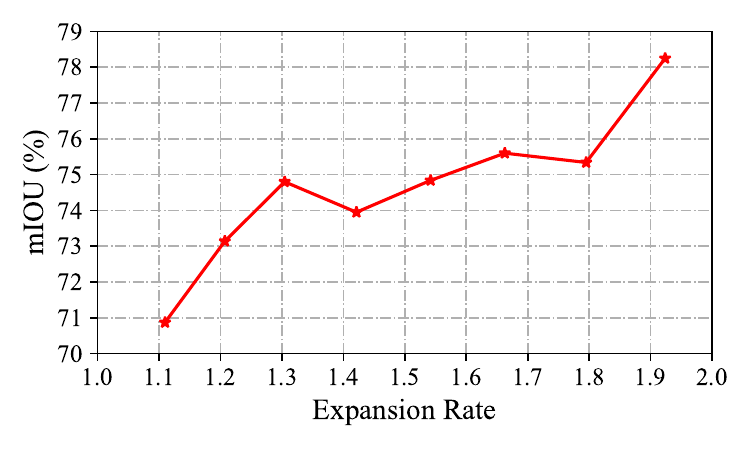}
	\end{minipage}
	\caption{The performance of the stego DNN model at different settings of the expension rate, left: intra-task steganography, right: inter-task steganography.}
	\label{fig:capacity}
\end{figure*}




\vspace{3mm}
\subsubsection{\textbf{Capacity}}
\label{sec422}
The capacity of the proposed method is evaluated in terms of the expansion rate defined as
\begin{equation}
e=\frac{N_{stego} }{N_{sec}}, 
\end{equation}
where $N_{sec}$ and $N_{stego}$ refer to the number of parameters in the secret and stego DNN model, respectively. Thus, we have $e \geq 1$, which is closely related to the number of interference filters that are inserted. As what we have mentioned before, we select top $30\%$ important insertion positions for filter insertion, the corresponding expansion rate is 1.52 and 1.60 for intra-task steganography and inter-task steganography, respectively.

We further evaluate how the performance of our proposed method would be changed when the expansion rate varies. Figure~\ref{fig:capacity} shows the performance of the stego DNN model at different expansion rates. It can be seen that, with the increase of the expansion rate, the performance of the stego DNN model gradually increases on both intra-task and inter-task steganography. This is because more parameters could be tuned for training the stego DNN model when the expansion rate increases. It could also been seen from 
\begin{table}[htbp]
	\centering
	\caption{The Undetectability of the proposed method.}
	\label{tab:Undetectability}
    \scalebox{0.95}{
		\begin{tabular}{c|c}
			\hline
			Detection method      & Detection rate (\%) \\
			\hline
			SVM & 52.50\\
			Backdoor detection \cite{xu2021detecting} & 50.00 \\
		     \hline
		\end{tabular}%
        }
\end{table}
Figure~\ref{fig:capacity} that our current expansion rate, which is around 1.6, ensures a good balance between capacity and fidelity for different steganography scenarios.

\vspace{3mm}
\subsubsection{\textbf{Undetectability}}~
To evaluate the undetectability, we  establish a model pool with different stego DNN models and clean DNN models trained. 
We select five benchmark datasets: MNIST \cite{lecun1998gradient}, Fashion-MNIST, GTSRB \cite{stallkamp2012man}, CIFAR10, CIFAR100 to generate $\mathrm{A}^2_5 = 20$ different secret-stego dataset pairs. 
Then, we train five well-known DNN models: LeNet-5 \cite{lecun1998gradient}, AlexNet \cite{krizhevsky2012imagenet}, VGG16 , ResNet18 and ResNet50 \cite{he2016deep} on the 20 dataset pairs to generate 100 different stego DNN models. Since we can only obtain 25 clean DNN models on the five datasets and five DNN models, we train each clean DNN model four times with different initialized parameters for augmentation. Totally, we obtain 100 stego DNN models and 100 clean DNN models in our model pool.

Training a classifier to detect the difference in the parameter distribution of stego DNN models and clean DNN models is a straightforward method. Following the idea, we compute a 100 bin-histogram from the parameters in each model to get a 100-dimensional feature vector for training a support vector machine (SVM) classifier. Specifically, we randomly select 80 pairs of such features from the stego and clean DNN models for training and the remaining 20 pairs are used for testing. The detection accuracy is shown in the first row of Table \ref{tab:Undetectability}, we can see that the SVM classifiers is not able to detect the existence of the secret DNN models from the stego DNN models with detection accuracy close to $50\%$. 


Additionally, we regard the secret DNN model as a backdoor\cite{gu2017badnets} of the stego DNN model and use the state-of-the-art DNN backdoor detection method \cite{xu2021detecting} to detect the existence of the secret DNN models. The detection accuracy is shown in the second row of Table \ref{tab:Undetectability}. It can be seen that the detection rate is 50\%, which is not better than random guess.  These results in Table \ref{tab:Undetectability} indicate the undetectability of the proposed method.



\begin{figure}[t]
	\centering
	\includegraphics[width=0.81\linewidth]{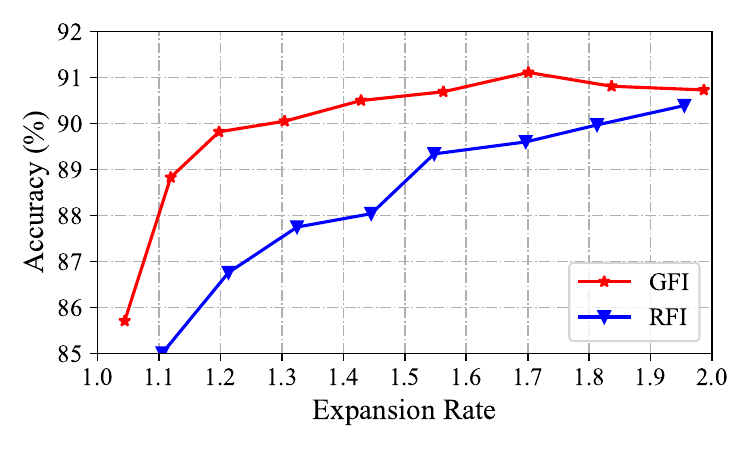}
	\caption{Comparisons between the gradient based filter insertion (GFI) and random filter insertion (RFI).}
	\label{fig:abs_gps}
\end{figure}

\begin{table}[t]
\renewcommand{\arraystretch}{1.1} 
  \centering
    \caption{Advantage of the proposed loss function $\mathcal{L}_{all}$.}
  \label{tab:abs_ld}
  \scalebox{0.70}{
  \resizebox{\linewidth}{!}{
\begin{tabular}{l|c|cc}
\hline
             Model         & Layer & $\mu$ & $\sigma$ \\ \hline
\multirow{2}{*}{$G_{\gamma}(\cdot)$} 

                            & \textrm{2}    & -0.03063 & 0.1731 \\
                            & \textrm{4}    & -0.02105 & 0.1777 \\
\hline
\multirow{2}{*}{$G_{\delta}(\cdot)$ with $\mathcal{L}_{all}$} 
                            & \textrm{2}    & -0.03064 & 0.1731 \\
                            & \textrm{4}    & -0.02047 & 0.1772 \\
\hline
\multirow{2}{*}{$G_{\delta}(\cdot)$ with $\mathcal{L}_{st}$}
                            & \textrm{2}    & -0.01700 & 0.1472 \\
                            & \textrm{4}    & -0.01128 & 0.1563 \\
                       
\hline
\end{tabular}}}
\vspace{-0.1cm}
\end{table}

\vspace{3mm}
\subsection{Ablation Study}
In this part, we conduct experiments on intra-task steganography to verify the effectiveness of our filter insertion strategy (i.e., GFI) and the loss function (i.e., $\mathcal{L}_{all}$) for model embedding.

\begin{table*}[t]
\renewcommand{\arraystretch}{1.1} 
  \centering
    \caption{Performance comparison among different schemes for embedding DNN models.}
  \label{tab:comparison}
  \resizebox{1.0\linewidth}{!}{
\begin{tabular}{c|c|c|c|c|c}
\hline
           \multirow{2}{*}{Category}         & \multirow{2}{*}{Method}  &Capacity   & Recoverabilit &  Undetectability & Performance reduction \\
                                    &              & Size of Stego/watermarked-data & BER(\%)/PSNR(dB)  & Eoob(\%)/detection rate(\%) & of stego DNN(\%) \\ 
           \hline
\multirow{2}{*}{DNN} &Uchida \textit{et al}. \cite{uchida2017embedding}& 142.94GB  &0/- & - & $\uparrow 0.07$ in ACC\\
\multirow{2}{*}{Watermark} &Fan \textit{et al}.\cite{fan2021deepip}              & 1.02TB &0/- & - & $\downarrow 1.68$ in ACC\\
                           &Guan \textit{et al}.\cite{guan2020reversible}        & 8.19 GB &0/- & - & $\uparrow 0.20$ in ACC\\
\hline

\multirow{2}{*}{Traditional} & Liao \textit{et al}. \cite{9124671}               &  160.00MB(0.50bpp)     &0/-    &33.30/-     & -\\
\multirow{2}{*}{Steganography}  &Tang \textit{et al}. \cite{tang2017automatic}   & 200.00MB(0.40bpp)    &0/-    & 16.20/-    & -\\
                           &Tang \textit{et al}. \cite{tang2020automatic}       & 160.00MB(0.50bpp)     &0/-    & 26.49/-    & -\\
\hline
                           &Zhu \textit{et al}. \cite{zhu2018hidden}            & 10.00MB(8.00bpp) &-/35.70 & -/76.49    & -\\
\multirow{3}{*}{DNN-Based} &Bakuja  \textit{et al}. \cite{Baluja2019image}      & 10.00MB(8.00bpp)      &-/34.13 & -/99.67    & -\\
\multirow{3}{*}{Steganography} &Weng  \textit{et al}. \cite{weng2019high}       &  10.00MB(8.00bpp)     &-/36.48 & -/75.03    & -\\
                           &Jing  \textit{et al}. \cite{jing2021hinet}          & 10.00MB(8.00bpp)      &-/46.78 & -/55.86    & -\\
                           \cline{2-6} 
                            &\textbf{Ours:intra-task}                     & 15.20MB($e=1.52$)     &0/-     & -/52.50 (SVM) and    & $\downarrow 0.93$ in ACC \\
                            &\textbf{Ours:inter-task}                     & 16.00MB($e=1.60$)     &0/-     & -/50.00 (Backdoor detection)    & $\downarrow 1.24$ in mIOU \\
                            
\hline
\end{tabular}}
\vspace{-0.1cm}
\end{table*}



\textbf{Effectiveness of GFI.}~We conduct random filter insertion (RFI) for model embedding instead of using the GFI. Figure \ref{fig:abs_gps} shows the classification accuracy of the stego DNN model at different expansion rates using different filter insertion strategies. We can observe that our GFI achieves better performance at all the expansion rates compared with using RFI. When the expansion rate is 1.2, the accuracy of the stego DNN model using GFI is over 3\% higher than that with RFI. At the accuracy of 90\%, the expansion rate of using GFI is around 1.3, which is significantly lower than using RFI (around 1.8). This demonstrates the effectiveness of our GFI for efficient model embedding.

\textbf{Effectiveness of $\mathcal{L}_{all}$.}~During model embedding, we train the stego DNN model using two different loss functions, including $\mathcal{L}_{all}$ and $\mathcal{L}_{st}$. Table \ref{tab:abs_ld} gives the mean and standard deviation of the first and second convolutional layers of the stego DNN model $G_{\delta(\cdot)}$ and the clean network $G_{\gamma}(\cdot)$. By using the $\mathcal{L}_{all}$ loss, the distribution of the parameters in the stego DNN model is much closer to the clean model compared with only using $\mathcal{L}_{st}$. This indicates that our proposed loss function is effective to construct a stego DNN model with a similar parameter distribution to the clean model.



\subsection{Comparisons}
In this section, we compare our proposed method with the existing state-of-the-art steganographic schemes and DNN watermarking schemes for hiding DNN models in terms of capacity, recoverability, undetectability and fidelity (for DNN watermarming schemes only). For fair comparison among different types of information hiding schemes, we assume the data to be hidden is a secret DNN model with size of 10MB for all the methods. The capacity is evaluated as the size of stego-data (or watermarked-data), the recoverability is reported as the BER or PSNR of the recovered secrets, the undetectability is indicated as the the “out-of-bag” error rate (EOOB) or detection accuracy (50\% means random guess), and the fidelity is indicated as the performance reduction of the stego/watermarked DNN model due to data embedding. 

Table \ref{tab:comparison} reports the comparisons among different schemes. It should be noted that, for the DNN watermarking schemes, the capacity is calculated and reported based on the maximum amount of secret data that can be embedded (i.e., fully embedded) in a VGG11 cover DNN model. The results of the recoverability and fidelity are duplicated from literature, where the DNN model is not fully embedded to achieve a good balance among capacity, recoverability and fidelity. For the traditional steganographic schemes \cite{9124671, tang2017automatic, tang2020automatic}, we duplicate the EOOB rate reported at a certain payload for undetectability from each paper, where the payload is the bit per pixel (bpp), i.e., how many bits can be embedded in a pixel. The EOOB is computed as the mean of false alarms rate and missed detection rate. For the DNN-based steganographic schemes \cite{Baluja2019image, zhu2018hidden, weng2019high, jing2021hinet}, we duplicate the PSNR (on imagenet) and detection accuracy at a payload of 8bpp from results reported in \cite{jing2021hinet}. Here, we assume the 10MB secret data is a natural image with meaningful content for these DNN steganographic schemes for hiding images or video frames.

It can be seen from Table \ref{tab:comparison} that our scheme requires significantly less storage for the stego-DNN model compared with the existing DNN watermarking schemes. And the size of our stego-data is less than or equal to 1/10 of size of the stego-data that are required using the existing traditional steganographic schemes. Compared with the existing DNN-based steganographic schemes \cite{zhu2018hidden, Baluja2019image, weng2019high, jing2021hinet}, which are not able to conduct lossless data extraction, our scheme achieves lossless recovery with BER of $0$ and better undetectability close to 50\%. In addition, all the existing steganographic schemes involve the change of data format between the secret-data (a DNN model) and the stego-data (images or videos).

\section{Conclusion}
\label{section:conclusion}
In this paper, we propose deep network steganography to tackle the problem of covert communication of DNN models. Our method is task oriented, which is able to disguise the secret-learning task (accomplished by a secret DNN model) into an ordinary learning task conducted on a stego DNN model. To efficiently construct the stego DNN model, we propose a gradient-based filter insertion scheme to insert some interference filters into the secret DNN model, where only the insertion positions with high importance are selected. While the importance of the insertion positions are measured by computing the gradients of the filters in the secret DNN model for the stego-learning task, we also propose a partial optimization strategy (POS) to train a stego DNN model with the original parameters (in the secret DNN model) frozen, where two statistical loss terms are designed and incorperated to improve the undetectability of the stego DNN model. As such, the secret DNN model can be recovered losslessly from the stego DNN model with high undetectability. Various experiments demonstrate the advantage of our network steganography for covert communication of DNN models. 

\newpage
\bibliographystyle{ACM-Reference-Format}
\bibliography{main}










\end{document}